\documentstyle[11pt,aaspp4]{article}
\input epsf

\lefthead{}
\righthead{}
\begin{document}
\title{Composition Mixing during Blue Straggler Formation and Evolution}

\author{Eric L. Sandquist, Michael Bolte, and Lars Hernquist\altaffilmark{1}}
\affil{Board of Studies in Astronomy and Astrophysics, University of
California, Santa Cruz, CA~95064; erics, bolte, lars@ucolick.org}
\altaffiltext{1}{Presidential Faculty Fellow}

\dates

\begin{abstract} 
We use smoothed-particle hydrodynamics to examine
differences between direct collisions of single stars and binary star
mergers in their roles as possible blue straggler star formation
mechanisms. From our simulations we find in all cases that core helium
in the progenitor stars is largely retained in the core of the
remnant, almost independent of the type of interaction or the
central concentration of the progenitor stars.

We have also modelled the subsequent evolution of the hydrostatic
remnants, including mass loss and energy input from the hydrodynamical
interaction. The combination of the hydrodynamical and hydrostatic
models enables us to predict that little mixing will occur during the
merger of two globular cluster stars of equal mass. Unmixed and
completely mixed models can both explain the luminosities of the brightest
BSSs by the merger of two turnoff mass stars. In
contrast to the results of Proctor Sills, Bailyn, \& Demarque (1995),
we find that neither completely mixed nor unmixed models can match the
absolute colors of observed blue stragglers in NGC 6397 at all
luminosity levels. We also find that the color distribution is
probably the crucial test for explanations of BSS formation --- if stellar
collisions or mergers are the correct mechanisms, a large fraction of
the lifetime of the straggler must be spent away from the main
sequence. This constraint appears to rule out the possibility of
completely mixed models. For NGC 6397, unmixed models predict blue
straggler lifetimes ranging from about 0.1 to 4 Gyr, while completely
mixed models predict a range from about 0.6 to 4 Gyr.  
\end{abstract}


\section{Introduction}\label{intro}

Blue straggler stars (BSSs) are objects in open and globular clusters
that are brighter and bluer than the cluster main-sequence turnoff
(MSTO), appearing to follow an extension of the main sequence.  Over
the years a number of explanations for their existence have been
considered, including an extended period of star formation in the
cluster, prolonged main-sequence lifetimes for some cluster stars as a
result of mixing, and the formation of objects more massive than the
current-day MSTO via the merger of close binary stars or through
the direct collision of two or more stars (see Stryker 1993 and
Bailyn 1995 for recent reviews).

With the discovery of eclipsing contact, semi-detached and detached
binaries among the BSS or main-sequence stars in several globular
clusters (Niss, J$\o$rgensen, \& Lausten 1978; Mateo et al. 1990; Hodder
et al. 1992; Kaluzny \& Krzeminski 1993; Yan \& Mateo 1994; Rubenstein
\& Bailyn 1996; Yan \& Reid 1996) and the discovery of populations of
bright BSS near the cores of very dense clusters (e.g. Auri\`{e}re,
Ortolani, \& Lauzeral 1990; Paresce et al. 1991, Lauzeral et al. 1992;
Auriere et al. 1990; Yanny et al. 1994a,b), there is growing consensus
that most BSS in globular clusters are in mass-transfer binary systems
or are the remnants of stellar collisions. We will refer to these two
possibilities as ``binary merging'' and ``collisional merging''
respectively.

If these two processes are the main mechanisms for BSS production, then
BSSs become interesting in a broader context than as footnotes to
our understanding of stellar evolution. The binding energy in
primoridal binary systems has long been invoked as a source of energy
in globulars to prevent or halt core collapse and/or
provide the energy for post-core-collapse expansion (e.g. Hut et al.,
1992). If signatures of binary mergers could be recognized and the
numbers and lifetimes of the remnants accurately determined, it would
be possible to place observational constraints on the input energy of
binary ``burning''. Similarly, if the remnants of star-star,
binary-star and binary-binary interactions could be unambiguously
identified and the lifetimes of the remnants calculated, then
observations of these objects in clusters with different stellar
densities and velocity dispersions would give observational constraints
on the stellar cross-sections for interaction.

Bailyn (1992) suggested that BSSs formed during binary mergers could be
distinguished from collisional remnants by differences in photometric
properties. Based on the Benz \& Hills (1987) hydrodynamical
calculations of colliding stars, Bailyn assumed that the remnants of
collisional mergers would be effectively fully mixed, with larger
helium envelope contents than the progenitor stars or than the remnant
of a binary merger of comparable total mass. BSSs formed in collisional
mergers would therefore be bluer than the extension of the cluster main
sequence, and BSSs formed in binary mergers would tend to be redder -- these
remnants look like evolved versions of higher-mass (than either of
the original binary components) stars.  For a given total mass,
collisional remnants would also be more luminous than binary merger
remnants.  Bailyn \& Pinsonneault (1995) suggested that the composition
profile differences between BSSs formed by the two processes
would also lead to different lifetimes and evolutionary tracks in the
color-magnitude diagram. This would lead to different luminosity
functions for populations of BSSs formed via the two paths.
Specifically, BSSs created via binary merger would be seen less often
at the bright end of the luminosity function as the higher core helium
content reduces the main sequence lifetime of these stars compared to
BSSs created by stellar collisions. 

The observational data appear to support the idea of BSS formation via
the two mechanisms and implicitly support the difference in
composition profiles of the two different types of remnants.  Bailyn \&
Pinsonneault (1995) interpret the BSS luminosity function of 47
Tucanae (NGC 104) as being consistent with production via collisions
of single stars.  The brightest BSS appear to form a separate class in
several other globular clusters.  In M3 (NGC 5272), the BSSs exhibit
a gap in their radial distribution (Ferraro et al.  1993), and
the innermost BSSs have a different luminosity function from the outer
group (Bailyn \& Pinsonneault 1995), consistent with a collisional
origin in the center and binary origin in the outskirts.  In the
low-density cluster NGC 5053, the 12 brightest BSSs are more centrally
concentrated than the 12 faintest (Nemec \& Cohen 1989). The
post-core-collapse globular cluster NGC 6397 sports seven BSSs that
are significantly brighter than the rest of the known BSSs (Lauzeral
et al. 1992; Rubenstein \& Bailyn 1996).  Because the bright BSS
population in NGC 6397 shows very clear differences from the faint
population in magnitude and in physical location in the cluster (6 out
of 8 of the BSSs within $15^{\prime\prime}$ of the core are bright;
Lauzeral et al.  1992), it is an ideal test of formation mechanisms
for the most massive BSSs.  Proctor Sills, Bailyn \& Demarque (1995)
interpret these stars as being the results of stellar collisions
during which the remnants are fully mixed, based solely on the
agreement in color with stellar evolution models.  Although one W UMa
contact binary has been identified outside the core, none of the
observed BSSs (which include the core BSSs) show variability
(Rubenstein \& Bailyn 1996). This provides weak support for the
collision hypothesis.

For interpreting these observations, the question of whether or not
BSSs are chemically mixed during their formation is therefore a key
one.  Many studies have followed the hydrodynamics of stellar
interactions, partly in an attempt to understand possible BSS creation
mechanisms.  Benz \& Hills (1987, 1992) carried out smoothed-particle
hydrodynamics (SPH) simulations of collisions of equal and unequal mass
polytropes of index $n = 3/2$. In their simulations, they looked at
particle mixing at low resolution, dividing the progenitor stars into
four equal mass bins at four radial positions, and examined the
positions of the particles in the remnant.  They found that head-on
collisions mixed the material less than did grazing impacts (which
result in maximum mass loss).  So, if two turnoff-mass stars (with
significant helium in their cores due to nuclear processing) collided,
the core helium would be more highly mixed throughout the remnant in a
grazing collision than in a head-on collision.  Similarly, for unequal
mass collisions, Benz \& Hills found that the more massive star in the
collision was thoroughly mixed, and that the less-massive star settled
to the core of the more massive.  Both results would imply a larger
envelope helium abundance, and a mixing of hydrogen into the core of
the remnant, indicating that BSSs evolve off of the main sequence in a
nuclear timescale.

Although the Benz \& Hills work has been widely used to interpret
BSSs, it is not clear that their models are entirely appropriate.  For
BSSs to be created above the level of the MSTO, they must be drawn
from stars near the turnoff mass. MSTO stars have high density
contrast (the ratio of central density to average density), and so are
better modelled by polytropes of index higher than $n=3/2$. Lombardi,
Rasio, \& Shapiro (1995) presented results for grazing collisions of
equal-mass $n = 3$ polytropes (with a polytropic equation of state
having $\Gamma = 5/3$ to best model the gas behavior). If the
progenitor stars have composition profiles like those of turnoff
stars, the resulting remnant has a helium profile as a function of $m
/ M_{tot}$ that is essentially identical to the progenitors. Thus, the
higher density (and gravity) of the stellar cores protects them
against mixing until they are able to merge. The higher core helium
abundance then implies that the remnant BSS would have a core
hydrogen-burning lifetime that is much shorter than the main sequence
lifetime of a zero-age star of equal mass.

There is another potential source of mixing that could significantly
influence the magnitude and color of a blue straggler. As Leonard \&
Livio (1995) theorize, the energy that is input into the envelope of
the remnant of a collisional or binary merger can cause the star to
swell to pre-main sequence proportions, which results in large-scale
convective mixing. The major questions in this case are: how much
energy is needed to induce the star to mix completely, and can typical
stellar interactions in globular clusters provide enough energy?

In this paper, we attempt to model the entire merger process in
detail.  We present hydrodynamical simulations using both $n=3/2$ and
$n=3$ polytropes in head-on collisions, grazing collisions, binary
collisions, and binary mergers in order to examine the effects of the
character of the stellar interaction on hydrodynamical mixing. We use
this information to create one-dimensional stellar models that
incorporate the mixing and energy inputs derived from the
hydrodynamical simulations, so that we can also determine the effects
of convective mixing. Throughout, we will concentrate on equal-mass
star collisions, which are particularly relevant to modelling the
brightest BSSs.  In \S 2, we discuss details of the hydrodynamic and
hydrostatic calculations for the collisional and binary merger
mechanisms. In \S 3 we present the results of these calculations: the
helium profiles and the evolutionary paths of the remnants after the
interaction. In \S 4 we look into the observational implications of
our calculations, and consider the evidence for these methods of
creating blue stragglers.

\section{Calculations}

\subsection{Hydrodynamics Simulations}

We have modelled various kinds of stellar interactions using smoothed
particle hydrodynamics, in particular utilizing the TREESPH code
(Hernquist \& Katz 1989). Our computations follow those of Goodman \&
Hernquist (1991), with slight differences. We refer the reader to
these references for details of the operation of the code. Here we
provide a short summary of points that will be needed in other
parts of the paper.

The hydrodynamical calculations reported in this paper all involve progenitor
stars modelled as polytropes of index $n = 3/2$ or $n = 3$.
For the $n = 3/2$ simulations carried out in the first phase of this
study, we chose to integrate the entropy equation. For a pure
polytrope, the specific entropy of any two parcels of gas are
identical, making it straightforward to set up the initial model. For
$n = 3/2$, we are immediately given a realistic adiabatic index for
the gas of $\Gamma = 5/3$.  However, for $n = 3$ polytropes with
$\Gamma = 5/3$, it is more convenient to create the initial model by
adjusting the energies of the initial particles (Lai, Rasio, \&
Shapiro 1993), and to use the thermal energy equation to evolve the
particles. We have followed this procedure for our $n = 3$ runs.

We have chosen not to model changes in the equation of state in the
stars resulting from composition changes. To first order this can be
accounted for using polytropes of higher index for stars
that have more core helium. The higher molecular weight of helium
compared to hydrogen reduces the pressure that can be exerted by the
gas, resulting in a more centrally concentrated structure.

For all of our two-star collisions, we used 3000 SPH particles in each
star. For the Goodman \& Hernquist (1991) four-star interactions we
have analyzed, 2048 particles were used in each star. A spherically
symmetric spline kernel (Monaghan \& Lattanzio 1985) was used to
smooth each particle (the density goes to zero at a
distance $2h$ from the particle center, where $h$ is the smoothing
length). The smoothing length of each particle was
allowed to vary adaptively with time in order to ensure interaction
with only the $N_{s}$ nearest neighbors (we chose $N_{s} = 38$).

We have used an artificial viscosity of the form (Monaghan 1992):
\[ {\bf a}_{i}^{visc} = - \sum_{j} m_{j} \Pi_{ij} \frac{1}{2}
\left[\nabla_{i} W({\bf r}_{i} - {\bf r}_{j}, h_{j}) + \nabla_{i}
W({\bf r}_{i} - {\bf r}_{j}, h_{i}) \right] , \]
where
\[ \Pi_{ij} = \frac{-\alpha \mu_{ij} \overline{c}_{ij} + \beta
\mu_{ij}^2}{\overline{\rho}_{ij}} , \]
and
\[ \mu_{ij} = \left\{ \begin{array}{ll}
			\frac{{\bf v}_{ij} \cdot {\bf r}_{ij}}{h_{ij}
(r_{ij}^{2} / h_{ij}^{2} + \eta^{2})} & \mbox{for} \: {\bf v}_{ij} \cdot
{\bf r}_{ij} < 0 , \\
			0 & \mbox{for} \: {\bf v}_{ij} \cdot
{\bf r}_{ij} \geq 0 . 
			\end{array} \right. \]
Also, $r_{ij} = \left|{\bf r}_{i} - {\bf r}_{j}\right|$, ${\bf v}_{ij}
= {\bf v}_{i} - {\bf v}_{j}$, $\overline{c}_{ij} = (c_{i} + c_{j}) /
2$ (the average sound speed for two particles), $h_{ij} =
(h_{i} + h_{j}) / 2$, and $\overline{\rho}_{ij} = (\rho_{i} +
\rho_{j}) / 2$. Here we have used $\alpha = 0.5, \beta = 1.0$, and
$\eta^{2} = 0.01$.

For the majority of our initial polytrope models, the particles were
distributed randomly in the star's volume, and were given masses
proportional to the density of the appropriate polytrope. (For runs
that used equal-mass particles, the positions of the particles were
weighted according to the local density of the polytrope.)  A star
created in this way is prone to oscillations because of small
irregularities in the sampling of the gas volume. We allowed the star
to adjust by introducing frictional damping for a short period of
time.

Our simulations of single star interactions generally fell into one of
three categories: (1) binary merger (BM) --- the two stars were
placed on circular orbits with synchronous rotation in order to
simulate a tidally-locked binary; (2) head-on collision (HC) --- the two
stars were given velocities in the center-of-mass frame that caused them
to collide with zero impact parameter; (3) grazing collision
(GC) --- the two stars were given velocities that caused them to collide
with an impact parameter $0 < R_{min} / (R_{1} + R_{2}) < 1$, where $R_{1}$ and
$R_{2}$ are the initial radii of the two stars. Because we are
only interested in interactions typical of globular clusters, we have
in all HCs and GCs used $V_{\infty} = 0$, where $V_{\infty}$ is the
relative velocity in the center-of-mass frame at infinite separation.
In addition, the stars in HCs and GCs were given initial separations
of $5(R_{1} + R_{2})$ in order to strike a compromise between
computational time and the initial gravitational perturbations on the
input stars. For the BMs of $n=3/2$ polytropes, we started the stars
with a separation $a = 1.3 (R_{1} + R_{2})$, and for the $n=3$ polytropes,
the separation was $a = 1.1 (R_{1} + R_{2})$. These separations were
large enough that the stars were able to fully relax in the mutual
potential before merging occurred.

The interactions were followed until a relaxed remnant was formed in
each case. This was judged by the lack of variation of the total
internal and gravitational energies with time during each run. In this
way we hope to have adequately modeled the hydrodynamical mixing that
occurs during the formation of the remnants. For HCs, the remnants
were nearly quiescent --- approximately spherical with extremely low
particle kinetic energies. For GCs and BMs though, the input angular
momentum typically created a remnant that was oblate, and in some
cases also had a disk of material. The spin-down timescale for these
kinds of remnants was too long to be modeled further by
hydrodynamical means. To determine the final mass of these remnants, we
followed the algorithm of Benz \& Hills (1987), and examined each
particle to see if it had enough energy to escape.  This was judged to
be true if $E > 0$ for the particle, where \[ E = E_{K} + P / (\Gamma
- 1) + \rho \Phi ,\] where P is the pressure on the particle, $\rho$
is particle density, $\Phi$ is the particle's gravitational potential,
and \[ E_{K} = \left\{ \begin{array}{ll}
			0 & \mbox{if} \: x v_{x}, \; y v_{y}, \;
\mbox{or} \: z v_{z} <
0, \\
			\frac{1}{2} \rho \left[ v_{x}^{2} + v_{y}^{2} +
v_{z}^{2} \right] & \mbox{otherwise}
		   \end{array} \right. \]
is the outgoing kinetic energy.

\begin{figure}[t]
\epsscale{0.7}
\hspace*{-1.5 true cm}
\plotone{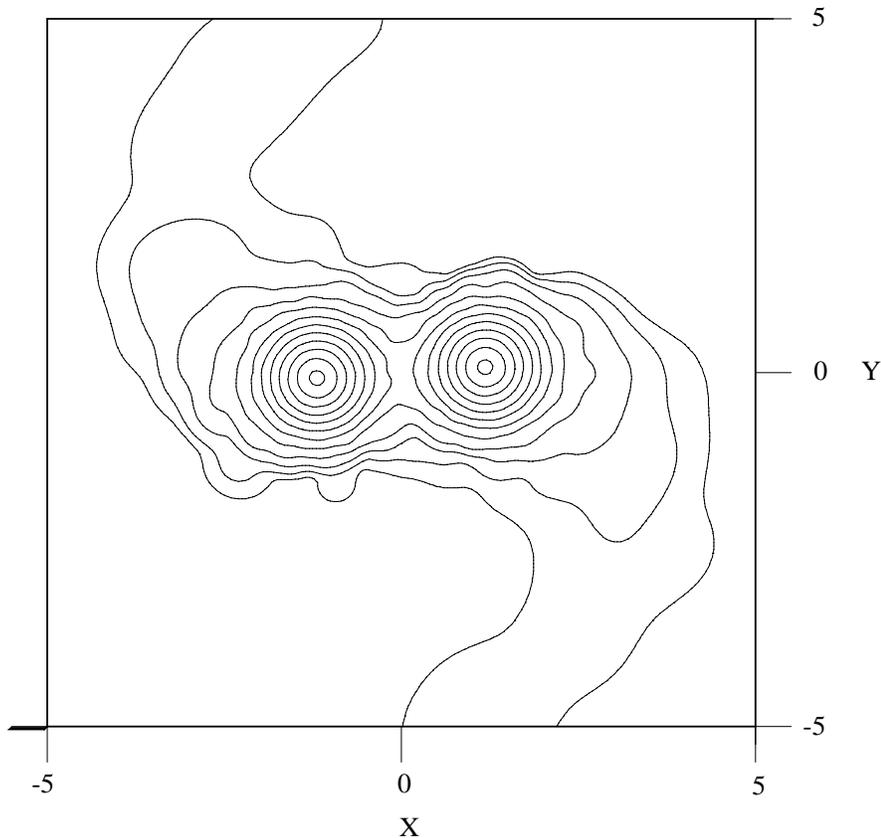}
\caption{Density contours for the binary merger of two $n=3$
polytropes. This represents the gas configuration before friction was
applied. The contours are evenly spaced two per decade, and range from
densities (in the orbital plane of the binary) of $10^{-5}$ g/cm$^{3}$
to 10 g/cm$^{3}$. \label{fig4}}
\end{figure}

We note that our BM run with polytropes of index $n = 3$ would not have
merged on a timescale that it was possible to model.  Once started, the
binary immediately formed a common envelope configuration that lost
energy at a very low rate even though the stars were started with a
small initial separation. This shows that dynamical instabilities in
synchronized binaries of the type discussed by Rasio \& Shapiro (1995)
occur for much smaller separations in $n = 3$ polytropes than they do
for $n=3/2$ (if they occur at all), or the timescale for the
instability is longer than we could reasonably simulate. We ended up
having to force the merger to occur by applying a small amount of
friction to the orbital motions. The density contours for the
common-envelope binary are shown in Figure~\ref{fig4} before friction
was applied.  This run should be regarded as somewhat suspect because
physical processes like convection could have time to occur if the
merger timescale is long enough.

\subsection{Hydrostatic Evolution Calculations}

To follow the subsequent phase of quiescent stellar evolution, we
carried out a number of one-dimensional stellar evolution
calculations. In order to create a realistic initial model, the
helium distribution in the remnant from the hydrodynamical
calculations had to be determined, and projected into a radial
profile. To accomplish this (for remnants that were oblate, or
had disks), we calculated the moment of inertia tensor for
the particles bound to the remnant. The remnants always had a high
degree of rotational symmetry, so we used the angular momentum vector
to define the minor axis. The moments of inertia about the three principal
axes were then used to calculate the oblateness of the remnant. For
axis lengths $a \geq b \geq c$ (and moments of inertia $I_{3} \leq
I_{2} \leq I_{1}$), the oblateness of the remnant was characterized by
\[ \frac{a}{c} \simeq \frac{a+b}{2c} = \frac{\sqrt{I_{1} - I_{2} +
I_{3}} + \sqrt{I_{1} + I_{2} - I_{3}}}{2 \sqrt{-I_{1} + I_{2} +
I_{3}}} .\]

The oblateness was typically computed for a small radial range near the
center of mass of the remnant, because there were definite changes as a
function of radius. It was most important to get the average oblateness
of the core of the remnant to ensure that the helium profile there was
computed as accurately as possible. Once this was done, the corrected
radius of each particle was calculated, and the particle's mass was put
into bins. The helium profile calculated from this procedure was then
mapped onto a pre-existing one-dimensional stellar model.  To test the
extreme case of complete mixing in BSSs, we computed the average helium
mass fraction from helium profiles derived in our SPH simulations, and
then imprinted the average helium abundance throughout.

These calculations were made with a version of the Eggleton stellar
evolution code (Eggleton 1991). The Eggleton code utilizes an adaptive
mesh, which is adjusted to put points where variations in mass,
temperature, and pressure are the largest. In our calculations, we have
followed the evolution of only $^{1}$H, $^{4}$He, $^{12}$C, and
$^{16}$O, as the error in the helium abundance introduced by the small
reaction network is easily dwarfed by uncertainties in the
hydrodynamical mixing, as we will discuss later.

We use the equation of state of Eggleton, Faulkner \& Flannery (1973).
More importantly, we use opacities for $\alpha$-enhanced
compositions calculated by the OPAL group (Rogers 1996; see Iglesias,
Rogers \& Wilson 1992 for a description of the physics). We have
chosen to use the most up-to-date opacities that are available
because they are known to primarily affect the color of a star,
which will be important later in comparisons with observational data.
For low temperature opacities, we have used the tables of Weiss,
Keady, \& Magee (1990). This choice is not particularly important
because surface temperatures for blue stragglers near the MS are
higher than the low temperature edge of the OPAL tables. As a result,
the position of our models in the HR diagram during the blue straggler
phase will not be affected.

An $\alpha$-element enhancement of $+0.3$ dex was also included as
globular clusters are known to have composition enhancements of this
order (Pilachowski, Olszewski \& Odell 1983; Gratton, Quarta \&
Ortolani 1986). These enhancements were taken into account explicitly
in the $^{16}$O abundances in the reaction network and in the OPAL
opacities.  Model luminosities and effective temperatures were
converted to the observational plane via bolometric corrections and
color transformations from VandenBerg (1992).  Table~1 lists data for
evolutionary runs from the main sequence for these runs.

As suggested by Leonard \& Livio (1995), the kinetic energy of the
stellar interaction can be converted into thermal energy, causing
the remnant to swell like a pre-main sequence star, potentially
inducing convective mixing in the remnant. To test this hypothesis, we
have computed models that start from a pre-main sequence
configuration. This has been accomplished by turning off convective
mixing of composition, and then gradually increasing the magnitude of
a constant energy generation (per unit mass) term for each model zone
until the star was placed well onto the Hayashi track. The
constant energy generation term could then be turned off, allowing
the star to contract back toward the main sequence on a thermal
timescale.

We do not attempt to model the evolution of the remnant between the
end of the SPH calculation and the beginning of this ``pre-main
sequence phase''. However, we believe that the details of this
adjustment are not particularly important. First, any convective
mixing will occur on timescales shorter than the
thermal adjustment timescale. As such, if convective mixing occurs
before the remnant becomes hydrostatic, it will almost certainly be
convective to about the same extent at the beginning of the
pre-main-sequence-like phase.  Mass loss or angular momentum loss from
the remnant could affect this argument, but we will examine it again
in \S~\ref{analy}.  Second, as we will see later, for typical BSS
masses, convection actually tends to be suppressed by the energy added
during the collision. No energy input was applied to the
completely-mixed models, as the thermal relaxation timescale is short
compared to the nuclear evolution timescale, so that the energy input
would not significantly change the expected BSS lifetime.

\section{Results}\label{res}

Table~2 summarizes the SPH calculations that we have made or
reanalyzed.  Column (2) gives the mass of the most massive remnant in
each interaction in units of the initial star mass. Column (3) gives the
oblateness of the core of the remnant (taken to be any particles with
a deprojected radius $r < 0.5$ in units of the initial star radius).
Columns (4) -- (6) give the kinetic, internal and gravitational energies
of the particles bound to the remnant (in units of $GM^{2} / R$, where
$M$ and $R$ refer to the input stars).  The table also lists energy
values for polytropes having $\Gamma = 5/3$. Finally, column (7) gives
the helium mass fraction at the center of the remnant, assuming that
the progenitor stars both had helium profiles matching that of a star
at the turnoff ($0.794 M_{\odot}$).

We have re-examined data from Goodman \& Hernquist (1991) for
collisions between equal-mass binary stars. These simulations involved
polytropes of index $n = 3/2$ with equal separations $a$, placed on
parabolic orbits which brought them to pericenter separations between
0 and $3a$.  Each star had a smaller number of particles than the ones
in our two-star collisions, but
tests indicate that this has little effect on the calculated helium
profile. These simulations should be useful for checking the effects
of multiple mergers, and for examining the differences introduced by
the altered kinematics of the interaction. Runs 22 and 43 from Goodman
\& Hernquist have been omitted from consideration because one star
involved in the merger had not completely merged into the remnant by
the end of the simulation. As a result, a bump appears in the helium
profile away from the core.

We have performed other simulations to test the effects of our
parameter choices in the SPH formalism. One such comparison can be
seen in the runs HCQ1 and HCQ1BH for $n = 3/2$. These simulations
tested the sensitivity of our choices relative to those of Benz \&
Hills (1987) --- specifically with respect to the form of the artificial
viscosity and the smoothing length for particles. With constant
smoothing lengths $h$ (in time) and $\beta = 0$ in the artificial
viscosity, we have been able to essentially reproduce the energy
evolution of Benz \& Hills's $R_{min}/(R_{1} + R_{2}) = 0.0$ and
$V/V_{esc} = 0.0$ run. We believe the form of the artificial viscosity
we use is more realistic (Hernquist \& Katz 1989), but it makes fairly
little difference. We do find more mixing of hydrogen into the core,
and more energy input into the remnant's envelope.

Modeling of $n = 3$ polytropes is difficult due to the large density
contrast between core and surface. As a result, we have completed runs
(HCQ1 and HCQ1EM) that test the effect of resolution on hydrodynamical
mixing and envelope energy input. In run HCQ1, the SPH particles were
given masses proportional to the density at their positions in the
input stars. In run HCQ1EM, all particles had equal mass, giving us
improved resolution of the core regions at the expense of less
accurate modeling of the outer envelope. The amount of mixing is
found to be larger in the HCQ1 run, and appears to be due to a small
number of massive (helium-rich) core particles in the initial stars
being pushed away from the core. The envelope energy input is also
slightly higher in the HCQ1 run, which is probably due to the improved
treatment of the initial contact between the two stars. We conclude
that our use of unequal mass particles probably means that our
estimates of envelope energy input are accurate, while our estimates
of hydrodynamical mixing are {\it overestimated} (see Lombardi, Rasio,
\& Shapiro 1995 for a short discussion).

\begin{figure}[p]
\plotone{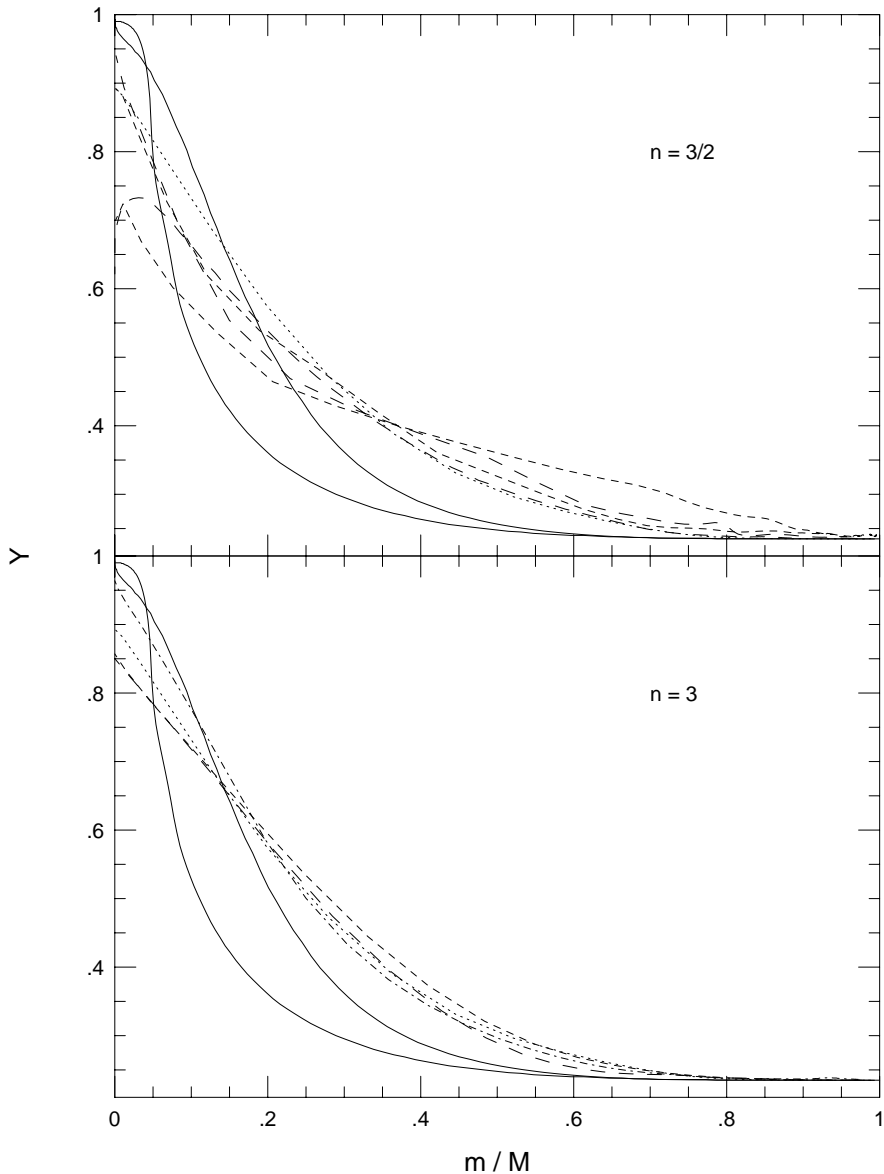}
\caption{Helium profiles for two-star mergers. The solid line
indicates the profile for the initial stars, the short dash lines show
the profiles for the most- and least-mixed of the mergers during binary
encounters (runs 54 and 57 respectively from Goodman \& Hernquist 1991),
the long dash line is an $n = \frac{3}{2}$ binary merger, and the dotted
line is an $n = \frac{3}{2}$ head-on collision, and the dot-dashed
line is an $n = 3$ grazing collision (the $n=3$ head-on collision is
virtually identical).   \label{fig1}}
\end{figure}

\begin{figure}[t]
\epsscale{0.8}
\hspace*{-3 true cm}
\plotone{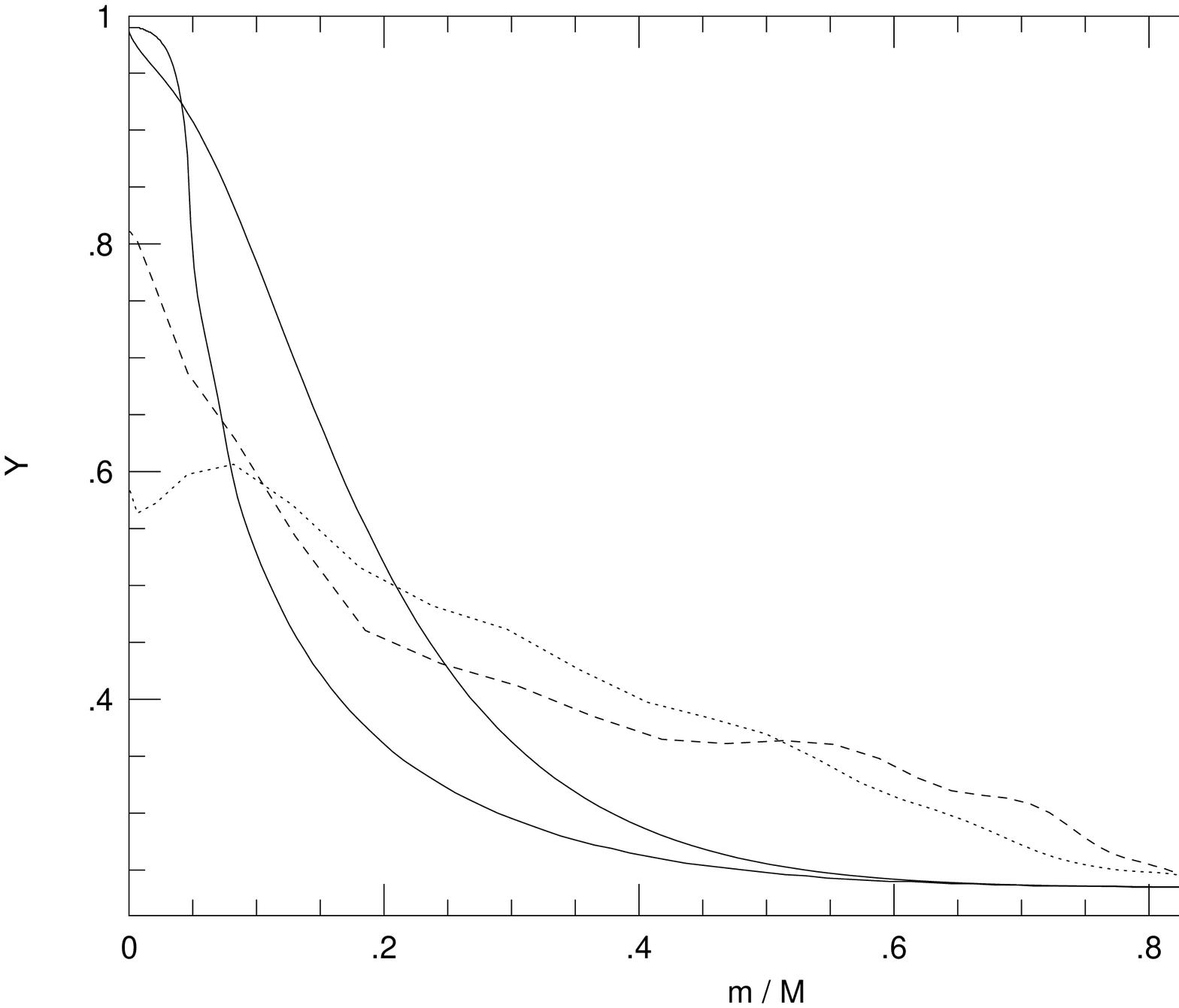}
\caption{Helium profiles for three-star mergers. The two profiles
shown are from simulations in Goodman \& Hernquist (1991): run 6 ({\it
dotted line}), and run 49 ({\it dashed line}).\label{fig2}}
\end{figure}

Figures~\ref{fig1} and \ref{fig2} show the helium profiles derived from
the simulations.  The input stars were given helium profiles
corresponding to a star near the cluster turnoff ($0.79 M_{\odot}$)
for a star with $Z = 0.00041$ (corresponding to an $\alpha$-enhanced
composition of [Fe/H] $= -1.91$) and $Y = 0.235$. We will discuss this
choice of composition in \S \ref{n6397}.


\begin{figure}[p]
\plotone{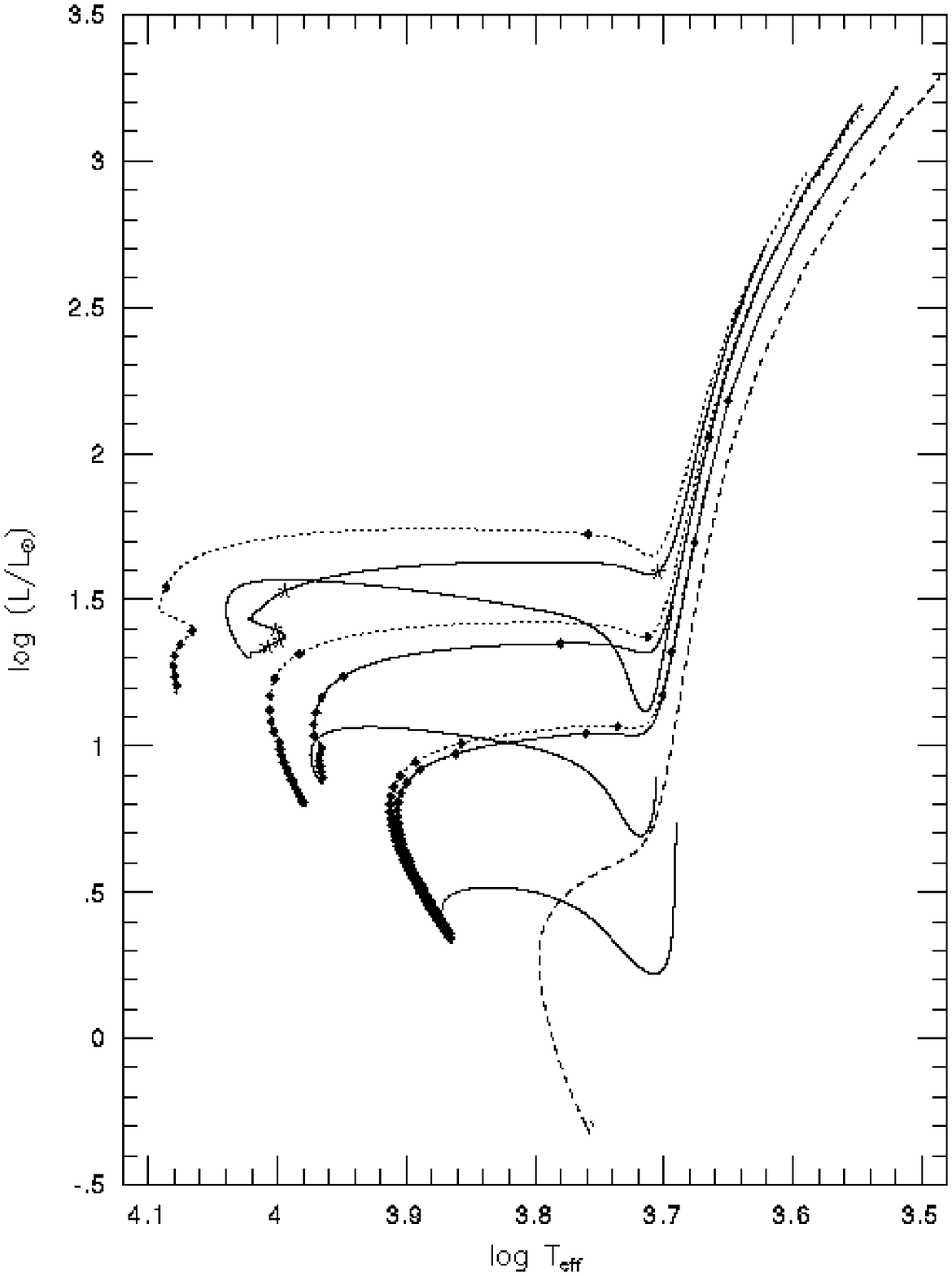}
\caption{HR diagram comparing our mixed ({\it dotted lines}) and
unmixed ({\it solid lines}) BSS models for masses of 1.47, 1.30, and
1.07 $M_{\odot}$ from most luminous to least luminous.  Symbols
on the evolutionary tracks indicate time intervals of $5 \times
10^{7}$ yr for the 1.47 $M_{\odot}$ unmixed model ({\it five-point stars}),
and $10^{8}$ yr for all other models ({\it filled circles}). A 0.79
$M_{\odot}$ track is also included ({\it dashed
line}), which has an age of 14 Gyr at its turnoff from the main
sequence. \label{fig3}}
\end{figure}

Table~3 presents the results of the hydrostatic evolution for the blue
straggler models. Figure~\ref{fig3} shows a comparison of some typical
evolutionary tracks for our BSS models in the HR diagram. We find that the
relaxation phase essentially follows the post-main sequence track to
the point where the remnant restarts stable burning of its remaining
core hydrogen. This is reasonable, given the similarity of pre-main
sequence and post-main sequence tracks near the zero-age main sequence.
This implies that collision or merger remnants that are relaxing back
to a hydrostatic configuration would be indistinguishable from
hydrostatic blue stragglers based on photometry alone. However, the
relaxation phase is short compared to the hydrogen-burning stage.

The larger the amount of hydrogen remaining in the core, the higher
the luminosity at which a hydrostatic structure is re-established on
the evolutionary track.  Based on our models, which predict little
mixing, we find that the majority of the early (long) MS phase is
avoided, considerably shortening the life of the BSS.

\section{Discussion}

\subsection{Model Analysis and Fits}\label{analy}

\subsubsection{Helium Profiles}

From the helium profiles shown in Figure~\ref{fig1}, it appears that
there is slightly more mixing in mergers involving $n = 3/2$ polytropes
than in $n=3$ polytropes in general. However, in every one of our SPH
simulations, we find that high helium abundances in the cores of the
progenitors are preserved during the interaction. This result is in
agreement with the simulation of an $n=3$ GC by Lombardi, Rasio, \&
Shapiro (1995). Our result is, however, in contradiction to the
``maximum mass-loss'' simulation of Benz \& Hills (1987) --- an $n=3/2$
GC. The histogram they present indicates much greater mixing of
helium-rich core material into the envelope.

Our simulations also show that the remnants of head-on collisions
are among the least-mixed of the runs. Grazing collisions and binary
mergers have slightly larger amounts of mixing, but have profiles that
are qualitatively similar to each other. {\it As a result, the
remnants of grazing collisions and binary mergers should evolve in
nearly identical ways.} If we are to judge from the hydrodynamical
mixing alone, neither fully-convective ($n \approx 3/2$; $M \leq 0.4
M_{\odot}$) nor radiative ($n \approx 3$; $M \geq 1.0 M_{\odot}$)
globular cluster stars will completely mix during a typical encounter.

\subsubsection{Convective Mixing}\label{conv}

Leonard \& Livio (1995) suggested that energy input from the
interaction could drive an expansion of the remnant's surface well
beyond what would be expected for a passively evolving main-sequence
star of the same mass.  For a large enough expansion, the star is driven
to the Hayashi track, where a surface convection zone develops that
mixes much of the interior. If this convection zone is extensive enough,
it could potentially mix core helium into the envelope. However, for
the highest mass BSSs, a couple of factors work against convective mixing
driven by this energy input.

First, energy input from the hydrodynamical interaction forces a
temperature decrease in the core as the gas expands.  As a result of
the reduced core temperature, the core temperature gradient is lowered,
which will {\it completely quench} core convection as long as thermal
energy is being released by the bloated remnant.  The more massive the
remnant is, the hotter it is as a main-sequence star. Consequently,
large energy inputs are necessary to force the star to expand to the
point where a surface convection zone can move inward to reach
helium-enriched layers.  The reverse of this process occurs in the
pre-main-sequence phase of evolution.

Second, we note that during pre-main-sequence contraction, it is the
core of the star that becomes radiative first. This occurs when the
central temperature is high enough that the matter is completely
ionized, and the density is high enough that electron scattering
is the dominant source of opacity. The resultant decrease in the
total opacity allows radiation to efficiently carry thermal energy
away from the core, eliminating the need for convection. A similar
effect occurs for BSSs driven to the Hayashi track, but to a greater
extent. Because of the high helium content of the core (mostly
unaffected by the hydrodynamical part of the merger, and also untouched by
core convection), the opacity is even lower than in a
pre-main-sequence star for the same physical conditions ($\kappa
\approx 0.20 (1 + X) \; \mbox{cm}^2 / \mbox{g}$ for electron scattering
in the Thomson limit). As a result, the higher core helium content can
prevent a surface convection zone from penetrating into the core {\it
even on the Hayashi track} if the remnant is massive enough.

The higher mass of the remnant also results in a different
distribution of convective zones when it returns to the main sequence.
For globular cluster ages of order 14 Gyr, turnoff stars have
masses of approximately $0.8 M_{\odot}$, with surface convection zones
containing relatively little mass. However, main-sequence stars with
mass greater than about $1 M_{\odot}$ develop core convection, while
the surface convection zone disappears.  Although convective cores
never achieve the extent necessary to completely mix the star in this
range of masses, they can mix material into the core from regions that
are richer in hydrogen. The higher the mass of the remnant, the larger
the extent of the convective core when the remnant has re-established
hydrogen burning via the CNO cycle. This effect can mix extra
hydrogen into the core. We find that for our most massive
model ($1.47 M_{\odot}$), the core convection zone covers the
innermost $0.2 M_{\odot}$, and increases the core hydrogen mass
fraction by about 0.1.  In contrast, core convection enriches the
core of our $1.30 M_{\odot}$ model by only 0.02 (covering $0.07 M_{\odot}$
of material), and core convection is never established in the $1.07
M_{\odot}$ remnant.

There are undoubtedly errors in the total energies of the SPH runs of
a few percent due to our use of adaptive smoothing lengths, and our
use of entropy equation integrations in the $n = 3/2$ runs (Hernquist
1993). However, the absolute accuracy of the total energies is not
important for this study because of the large amount of energy
necessary to drive the star back up the Hayashi track. The
relatively small kinetic energies involved in collisions in globular
clusters makes it unlikely that this energy input completely mixes the
star except in the rarest of situations.

We have seen in the discussion so far that the remnant mass has a
variety of implcations for the effectiveness of convection in the merger
remnants. As a result, we expect that mass loss during the merger
and later during spin-down will have some influence on the next
stages of the remnant's evolution. Hydrodynamical mass loss does not
exceed about 7\% in any of our simulations, which will not affect our
conclusions. A merger can also produce a disk around the remnant, which
could in turn be driven away via interaction with the central object
and angular momentum transfer. However, disks are formed in few of the
simulations, and even then the majority of the mass resides in
the central object. So, we believe that mass loss during the merger
process will not change our conclusions on the matter of mixing.

\subsection{NGC 6397}\label{n6397}

NGC 6397 is one of the nearest globular clusters, and photometry of
its BSSs has been carried out even into its dense core (Lauzeral
et al. 1992; Rubenstein \& Bailyn 1996).  The core population of BSSs
is significantly brighter than the majority of BSSs farther from the
center. The high central density of the cluster may indicate that
these stragglers are the result of stellar mergers. For these reasons,
this cluster makes an excellent subject for comparison between our
models and observations. Before proceeding with fits to the BSSs in
NGC 6397, we briefly discuss the characteristics of the cluster that
could influence the placement of the BSSs in a diagram of $M_{V}$
versus $(B-V)_{0}$.

We have chosen to use the Zinn \& West (1984) metallicity [Fe/H]
$= -1.91$. As summarized in Table~5 of that paper, the majority of
other studies have found values slightly higher: [Fe/H]
$\approx -1.8$. Some exceptions have been [Fe/H] $= -2.01$ (Smith
1984), $-2.24$ (Pilachowski 1984), and $-1.96$ (from one star;
Lambert, McWilliam, \& Smith 1992). Recently however, it has been
found that the Zinn \& West scale is decidedly nonlinear with respect
to a consistently reduced set of high-dispersion spectroscopic
determinations of cluster metallicities (Carretta \& Gratton 1996b).
If the Zinn \& West value is corrected using the formula given by
Carretta \& Gratton, we derive [Fe/H] $= -1.69$. Carretta \& Gratton
(1996a) measure a value [Fe/H] $= -1.82 \pm 0.04$. As a result
we will consider the potential metallicity error to be $\pm 0.2$ dex,
with somewhat higher probability going to higher metallicities.

We have also chosen to include enhancements of $\alpha$ elements in
order to derive the value for Z used in our models. Although there
have not been studies of these elements in NGC 6397, other globular
clusters are known to posses enhancements of approximately
[$\alpha$/Fe] $\sim +0.3$. On the whole, such enhancements make the
models redder and fainter, almost exactly mimicking an increase in
[Fe/H] (Chieffi, Straniero, \& Salaris 1991).

Determinations of the reddening of the cluster can be encompassed with
the value E$(B-V)= 0.18 \pm 0.02$ (Peterson 1993).  We will use the
Lauzeral, Auri\`{e}re, \& Coupinot (1993) distance modulus $(m - M)_{V}
= 12.50$, derived using the level of the horizontal branch (HB) and the
$M_{V}$(HB) -- [Fe/H] relation of Lee, Demarque, \& Zinn (1990). There
are two potential sources of error in this approach --- the zero-point
of the $M_{V}$(HB) -- [Fe/H] is still uncertain, and the blue HB of NGC
6397 makes determination of the level of the HB difficult.

\begin{figure}[t]
\plotone{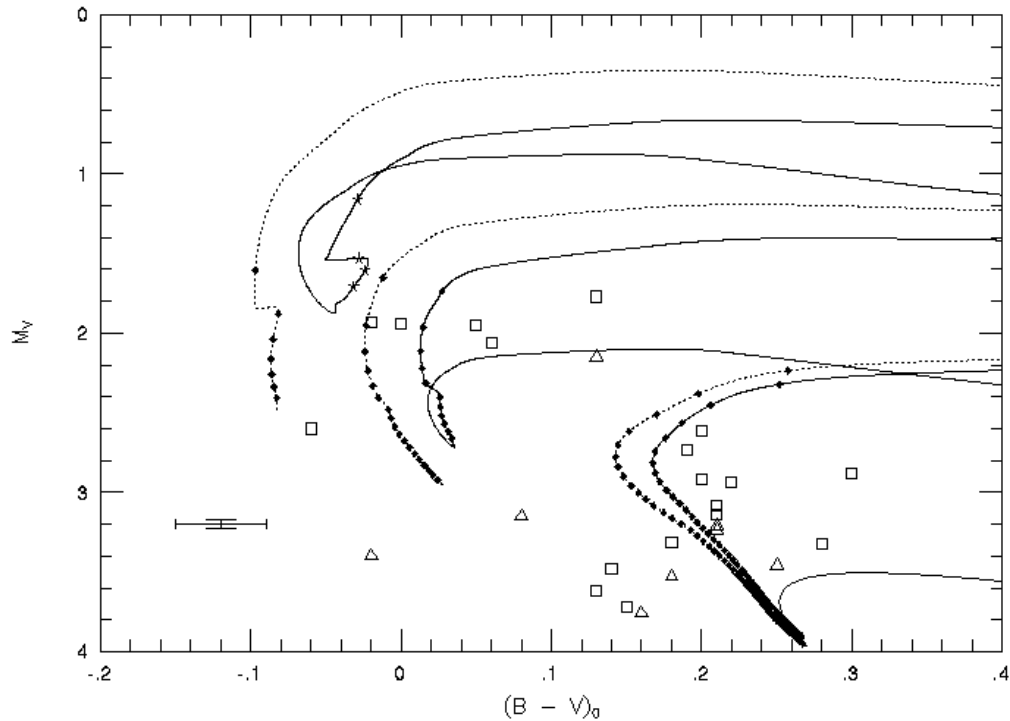}
\caption{Color-magnitude diagram comparing our mixed and unmixed
BSS models with observed BSSs in NGC 6397. The evolutionary tracks and
filled symbols have the same meaning as in Figure~\ref{fig3}. The blue
straggler photometry is from Lauzeral et al. 1992 ($\Box$) and from
Rubenstein \& Bailyn 1996 ($\triangle$). The error bars indicate the
maximum photometric errors for the BSSs quoted by Lauzeral et al.
1992.\label{fig5}}
\end{figure}

With this choice of distance modulus, reddening, and metal content,
along with an age of 14 Gyr, we can reproduce the turnoff magnitude
and color to good accuracy with a $0.794 \; M_{\odot}$ model:
$(B-V)_{TO} = 0.59$ and $V_{TO} = 16.6$, compared to 0.6 and 16.5 from
the cluster CMD (Lauzeral et al. 1992). Figure~\ref{fig5} shows a
comparison of our evolutionary tracks with the positions of NGC 6397
blue stragglers for this choice of parameters. We note that fitting
metal-poor subdwarfs to the main sequence yields a different distance
modulus $(m-M)_{V} = 12.1$ (Anthony-Twarog, Twarog, \& Suntzeff 1992).

\subsubsection{Luminosity}

A comparison of our evolutionary tracks with the photometry of BSSs in
NGC 6397 is shown in Figure~\ref{fig5}.  Because of the overall helium
enrichment in the remnants from prior evolution of the input stars, the
luminosities of both the mixed and unmixed BSS models are higher than
a zero-age star of the same mass.  During the core hydrogen-burning
phase, the luminosity of the unmixed model exceeds that of the mixed
model, primarily because the early stages of core burning are
avoided in the unmixed models. In any case, both sets of models can
produce remnants of two-star collisions that have sufficient
luminosity to explain the NGC 6397 observations.

The peak luminosities of our most massive BSS models ($\sim 2 M_{TO}$)
are larger than the brightest BSSs by nearly a magnitude. This could be
explained if BSSs are not produced with this high a mass, or if the
evolutionary timescales for high mass BSSs are very short compared to
stragglers of lower mass. While we cannot directly test this first
possibility, our BSS models show that our most massive unmixed model
has a lifetime about 21\% of the next most massive (a 12\% decrease in
mass). This may explain the lack of brighter blue stragglers.

If the no-mixing hypothesis is correct, then it may be necessary to
invoke mass loss in excess of what is found in the hydrodynamical
simulations. This could occur during spin-down of the merger remnant,
possibly via a magnetic braking mechanism (Leonard \& Livio 1995), even
though our models never become fully convective. The energy input from
the collision could cause the star to expand enough to have a
significant convection zone (in mass and radius) without going deep
enough to reach helium-enriched portions of the core. It is not
currently known if BSSs in globular clusters are rotating rapidly or
not, but BSSs in the open cluster M67 are known to have moderate
rotation rates (between about 20 and 120 km/s; Peterson, Carney, \&
Latham 1984; Mathys 1991).

\subsubsection{Absolute Colors}

The evolutionary tracks for our most massive unmixed BSS models cover
the colors of all but a few of the bluest stars. For the bright BSSs,
the completely mixed models can cover even the bluest BSSs. However,
for the faint BSSs though, neither set of models goes blue enough.
Given that the total systematic uncertainties quoted by Lauzeral et al.
(1993) are 0.1 mag in $(B-V)$, this color disagreement is not yet
a serious problem. If the colors are accurately determined though, this
must be explained.

Mass loss in addition to that resulting from the initial collision
would further increase the disagreement. The metallicity we have used
is probably at the low end of the acceptable range, so in all
likelihood a change in the value of [Fe/H] for the cluster would also
increase the disagreement. An abnormally low value of [$\alpha$/Fe]
relative to other clusters could improve the color agreement, but this
is unlikely.

The opacity data used in our models could also potentially affect the
absolute colors. From comparisons that have been made between OPAL and
Opacity Project datasets (Seaton et al. 1994), the opacities
calculated using two different approaches agree to high accuracy. In
addition, our comparisons of surface opacities (OPAL; Weiss et al.
1990; Alexander 1996) show that in the region of overlap that matters
to our models, the agreement is also very good.  For the surface
temperatures of BSSs (relatively near the main sequence),
uncertainties in surface opacities do not have an effect since
molecular opacity, which is omitted in the OPAL calculations, does not
become important until the star approaches the Hayashi track. We
conclude that the opacities themselves are unlikely to explain
the color mismatch.

\subsubsection{Color Distribution}

A quick examination of Figure~\ref{fig5} indicates that the models as
they stand predict that BSSs should populate a smaller range of colors
than is observed. In the past, this fact has not received much
attention for several reasons. Photometric measurement and calibration
errors both contribute to the scatter among the observed BSSs. In
Figure~\ref{fig5}, we have indicated the {\it random} errors in the
BSS photometry of Lauzeral et al. (1992) to show that this source
cannot explain the extra color scatter. In addition, observations have
shown that there are other ways of producing objects that populate the
regions where BSSs are found. Both detached and contact binaries are
found among the BSS populations in NGC 4372 and NGC 5466. These
examples show that stars that are in the process of merging can
also be observed as blue stragglers. Our picture of merger remnants is
undoubtedly confused by these stars in transit.

However, even among blue stragglers that are known to be single stars,
the color scatter is real.  SX Phe variables have been
found in $\omega$ Cen (J$\o$rgensen 1982; J$\o$rgensen \& Hansen 1984),
NGC 5466 (Mateo et al.  1988), NGC 5053 (Nemec et al. 1995) and NGC
4372 (Kaluzny \& Krzeminski 1993). The color distribution of samples
of these stars provides a more definite indication of the color
scatter, both because the pulsation mechanism can only operate in
relatively relaxed single stars, and because these stars must be
measured many times to derive light curves that show their identities.

In the globular cluster NGC 4372, 3 known SX Phe variables are
significantly redder by about 0.2 in $(B-V)$ than the remaining 5,
which may still form a fairly tight sequence (Kaluzny \& Krzeminski
1993). One of the 5 known SX Phe variables in NGC 5466 (Nemec et al.
1995) is bluer than the others by about 0.15 in $(B-V$). The apparent
bifurcation in color for these variables may be a result of different
modes of pulsation. However, it does indicate significant color
scatter among {\it single} stars.

Because stars evolve almost entirely to the red once core hydrogen has
been exhausted, the apparent color width of the BSS sequence can
indicate the amount of time a BSS spends completing core hydrogen
burning compared to the time it spends as a subgiant. As can be seen
from the evolutionary tracks shown in Figure~\ref{fig5}, a
completely-mixed remnant has a longer main-sequence lifetime, and
spends a larger fraction of its remaining life close to a
helium-enriched main sequence, prior to its turnoff from the main
sequence. So, if massive BSSs were somehow completely mixed in all
cases, they would define a comparatively thin extension of the cluster
main sequence.

Our models, which show little mixing, are able to better explain the
color width of the BSS sequence. If the high helium content of the core
of a merger remnant is preserved, the main sequence phase is truncated,
allowing the early subgiant branch phase of evolution to span a larger
fraction of the remnant's life. In addition, the pre-main-sequence-like
phase of evolution follows the post-main-sequence evolution reasonably
closely, slightly increasing the possibility that a remnant could be
observed in the subgiant region of the CMD.  But with the amount of
mixing we find in our models, too much time is spent relatively close
to the main sequence. This could be explained if the hydrodynamical
mixing is overestimated in our simulations. This is possible,
considering the arguments given in \S \ref{res}. It is also conceivable
that one or both of the stars involved in a merger had evolved past the
cluster turnoff, and hence has a higher core helium content to start.
Due to their increased sizes, such stars have larger geometrical cross
sections for collisions, and make binary mergers more likely due to
increased tidal interactions.

As one looks fainter, the progenitor stars must become less
massive. As a result, the cores of the remnant BSSs have lower
helium contents. This means that unmixed models spend an increasing
amount of time near the zero-age main sequence as the mass of the
remnant decreases, which would lead to a definite concentration of
stragglers toward the blue side of the region. As can be seen in
Figure~\ref{fig3}, the mixed and unmixed stars become more and more
alike as BSS luminosity decreases. Unequal mass collisions would not
help solve the problem --- if the remnant is completely mixed, the lower
mass star still contributes little helium, leading to a dilution; if
unmixed, the lower entropy material of the low mass star would
displace the relatively high entropy (and high helium content) matter
at the core of the high mass star. In both cases, the core of the star
still has a large amount of hydrogen, and will spend much of its life
near the main sequence. An additional mechanism appears to be
necessary to explain the color scatter at low BSS luminosity.

For the stars in NGC 6397 (and other clusters; Fusi Pecci et al.
1992), the scatter in the BSS colors appears to be inconsistent with
both hypotheses as they stand, even if photometric scatter (0.03 mag
in $V$ or $B-V$ at maximum; Lauzeral et al. 1992) is accounted for.
Differential reddening is potentially more important for NGC 6397
because of its relatively large average reddening (E$(B-V) = 0.18$;
Peterson 1993). However, small color widths of the blue HB and of the
RGB argue against this being a significant effect --- variations of
approximately 0.15 in E$(B-V)$ in the innermost $15^{\prime\prime}$ of
the core would be necessary to make the brightest BSSs consistent with
such a thin sequence.  Further, the color scatter observed in the BSSs
of many other clusters with much smaller reddenings cannot be
explained this way.

For the distribution of blue stragglers to be explained by the
evolution of stars more massive than the turnoff of the cluster, it
seems that at least two things must occur. First, there must be a
relatively large enrichment of helium in the envelope of the remnant
star, which is required to explain the bluest of the stragglers.
Second, there must also be nearly complete exhaustion of hydrogen in
the core of the remnant, as required by the seemingly uniform color
distribution of the stragglers in many clusters.

\subsection{Advanced Evolutionary States}

From our evolutionary tracks, we can predict regions in the
color-magnitude diagram (CMD) where BSS progeny might be found. These
topics have been discussed in other papers by Fusi Pecci et al. (1992)
and Taam \& Lin (1992). As red giants, evolved BSSs would be bluer
than undisturbed cluster giants.  As such, they would contribute to
the color dispersion on the RGB (Taam \& Lin 1992). Also, the most massive BSSs
undergo helium flash at lower luminosities in our models than other stars.
However, the RGB phase does not last long enough to be well-populated
in cluster CMDs, and field stars also add to the confusion in the
photometry, making it unlikely that evolved BSSs would be found on the
RGB. Fusi Pecci et al. (1992) argue that clusters with
the largest populations of BSSs will have an observable number of
stars on the reddest end of the HB as a result of their higher mass
relative to other cluster stars. Taam \& Lin suggest that anomalous
Cepheids may also be BSS progeny.

The mixing of helium into the stellar envelope will also affect the
position of the HB stars in the CMD. In general, higher envelope helium
abundance reduces the opacity, resulting in more compact, bluer stars.
For massive HB stars though, increased helium abundance primarily
increases the luminosity, but does not affect the color (Dorman 1992a). The
complete mixing hypothesis would predict that the progeny of the most
massive BSSs would have rather large envelope helium abundances, which
would make them significantly more luminous than the rest of the HB,
even to the point of looking like the faintest AGB stars.

More modest envelope enrichments occur in our unmixed models.
The amount of enrichment is rather small (see \S \ref{conv}), although
slightly dependent on the energetics of the merger process, as higher
energies imply slightly larger hydrodynamical mixing and, more
importantly, deeper penetration of the convective envelope into
helium-enriched regions during the pre-main-sequence-like phase.

From the models of Dorman (1992b), we can estimate the horizontal
branch lifetime for massive BSS progeny. Assuming that there is mass
loss between the RGB and HB of approximately $0.2 M_{\odot}$ (which
could produce the HB morphology of NGC 6397), an RGB mass of $1.5
M_{\odot}$ could be reduced to $1.3 M_{\odot}$. For an HB mass of $1.2
M_{\odot}$, Dorman's models predict that the core helium-burning
lifetime is somewhat less than 90 Myr.

The two mixing hypotheses make different predictions for the relative
number of BSSs and HB stars that are BSS progeny. Table~3 lists the
ages of several points in the evolution of our BSS models --- the start
point, the first relaxed model (when the star begins a net
absorption of thermal energy, as opposed to the thermal energy release
during contraction), and the turnoff (bluest) model. The complete-mixing
hypothesis predicts a BSS lifetime of approximately 0.65 Gyr for the
most massive remnant. Our most massive model for the no-mixing
hypothesis has a hydrogen burning lifetime of about 0.16 Gyr. Thus,
unmixed models would predict approximately equal numbers of massive BSSs and
their progeny, while completely-mixed models predict about 7 times as
many massive BSSs as progeny.

The agreement of this ratio for completely-mixed models with the value
of 6.6 from observations of globular clusters compiled by Fusi Pecci
et al. (1992) should not be taken as evidence for that scenario,
however.  Because lower mass BSSs must also be present (inheriting
lower core helium abundances, and thus longer lifetimes), unmixed
models bracket the BSS lifetime predicted from HB models and observed
BSS to BSS progeny ratios.  Completely-mixed models have lifetimes
that are in general too long.

\section{Conclusions}

1) From the results of our hydrodynamical simulations, we find that
significant mixing from core to envelope does not occur in stellar interactions
involving equal-mass stars, regardless of whether the interaction is a
head-on or a grazing collision, or a binary merger. Grazing
collisions and binary mergers result in slightly more mixing of helium
into the envelope than do head-on collisions.  (This result conflicts
with the findings of Benz \& Hills 1987.) Grazing collisions and binary
mergers produce remnants that appear to be extremely similar.

Stars with lower central concentration (less dense cores) suffer
slightly more hydrodynamical mixing, but not enough to alter the
subsequent evolution of the remnant. Our analysis of the Goodman \&
Hernquist (1991) simulations of binary collisions indicate that these
yield somewhat more hydrodynamical mixing than do single star
collisions.  As was found by Lombardi, Rasio, \& Shapiro (1995)
though, additional tests lead us to conclude that hydrodynamical
mixing in our simulations is likely to be overestimated due to the
finite spatial resolution of the gas in the model stars and the
artificial diffusion that results.

2) Mass loss in our hydrodynamical simulations never exceeds 7\% of the
total mass of the merging stars, and so is insufficient to
qualitatively alter our conclusions. Disks formed by stars colliding
with the largest impact parameters can contain a significant amount of
the total bound mass. In our analysis, we cannot estimate the amount of
disk material that will be lost through angular momentum transfer
during spin-down of the remnant. However, mass loss from a disk would
only tend to increase the disagreement between our models and the
observed colors of the hottest blue stragglers.

3) We have modeled the evolution of the merger remnants by imprinting
helium profiles from hydrodynamical simulations onto one-dimensional
stellar models. We have included effects of mass loss and energy input
from the collision.

Typical energy inputs from the collisions are unable to provide the
remnant with sufficient thermal energy to drive a convection zone large
enough to mix core helium into the envelope. The energy input also
temporarily quenches core convection in the more massive remnants by
reducing the core temperature. The higher helium content of the core
results in lower opacities, further preventing convective mixing of
helium into the envelope even if the star does reach the Hayashi
track.  Additionally, thermally-relaxing remnants tend to reside
in regions of the CMD populated by hydrostatic BSSs, making purely
photometric separation impossible.

4) Unmixed models predict a lower limit for BSS lifetimes in NGC 6397
of about 0.1 Gyr for the most massive BSSs, while completely mixed
models predict a minimum lifetime of about 0.6 Gyr.

For the most massive BSSs, completely mixed models can be as blue as
the bluest BSSs, and the luminosity of the core hydrogen burning phase
matches that of the brightest BSS in NGC 6397. However, these models
{\it cannot} explain the reddest BSSs or the color scatter in the
stragglers because they spend a large fraction of their time in the
core hydrogen burning phase.  The most massive unmixed models are more
luminous than the brightest BSS during their main sequence phase, but
also evolve rather quickly away from the main sequence which may make
them less observable. Slightly lower mass models can reproduce the
luminosity of the observed BSSs, but still have difficulty explaining
the bluest BSSs, and the color scatter.

For faint BSS populations, the completely-mixed and unmixed
evolutionary tracks have similar appearance because of the small
amounts of core helium present in the progenitor stars. Both sets of
models predict that the BSSs would inhabit a relatively large color
range in the CMD, although with a concentration of stars on the blue
side of the strip. Neither hypothesis is easily able to explain the
bluest BSSs. This is in contradiction to the results of Proctor Sills,
Bailyn, \& Demarque (1995) because their study did not account for
abundance enhancements in $\alpha$ elements that are known to occur in
globular clusters.

The color width of the BSS sequences in many, if not all, globular
clusters indicate that large portions of the lifetime of a blue
straggler must be spent away from the zero-age main sequence.
Completely mixed models cannot explain this observation because the
star spends a large amount of time with comparatively high
hydrogen content in the core, and hence lower luminosity and longer
nuclear burning timescale. Unmixed models can potentially explain this
observation because the vast majority of the core hydrogen-burning phase is
avoided. Observations of SX Phe pulsating variables (which are presumably
relatively relaxed single stars) in the clusters NGC 4372
and NGC 5466 indicate that there is significant color scatter among
single star BSSs. 

We must conclude that neither the completely mixed nor the unmixed
hypothesis is able to consistently reproduce all details of the BSS
distribution for NGC 6397, and that this would probably also apply to
most other clusters. However, the main prediction of the complete
mixing hypothesis --- that blue stragglers produced by stellar
collisions or binary mergers should populate a relatively thin locus in
the CMD --- appears to be untenable.

\acknowledgments

We would like to thank P. Bodenheimer, M. Davies, and S. Siggurdson for
helpful conversations, and F. Rogers and D. Alexander for providing us
with opacity tables for $\alpha$-enhanced compositions. This work was
supported in part by the Pittsburgh Supercomputing Center, the National
Center for Supercomputing Applications (Illinois), the San Diego
Supercomputing Center, and the NSF under grants AST-9420204 (M.B.),
ASC-9318185 (L.H.), and the Presidential Faculty Fellows Program.

\newpage

\begin{deluxetable}{ccccccc} 
\tablecolumns{7}
\tablewidth{0pc}
\tablenum{1}
\tablecaption{Results of Main Sequence Stellar Evolution Calculations} 
\tablehead{\colhead{$M/M_{\odot}$} & \colhead{Stage} & \colhead{Age (Gyr)} &
\colhead{$E_{I}$\tablenotemark{a}} &
\colhead{$E_{G}$\tablenotemark{a}} & \colhead{$\log R$ (cm)} & \colhead{$n_{eff}$}}
\startdata 
1.588 & ZAMS & 0.0 & 7.374 & $-14.90$ & 10.844 & 3.06 \nl
 & TO & 1.1 & 7.377 & $-14.85$ & 10.947 & 3.47 \nl
1.200 & ZAMS & 0.0 & 4.581 & \phn$-9.25$ & 10.821 & 3.12 \nl
 & TO & 3.1 & 4.867 & \phn$-9.79$ & 10.940 & 3.69 \nl
0.794 & ZAMS & 0.0 & 2.23 & \phn$-4.50$ & 10.692 & 2.75 \nl
 & TO & 14.0 & 2.59 & \phn$-5.20$ & 10.970 & 3.79 \nl
0.700 & ZAMS & 0.0 & 1.79 & \phn$-3.62$ & 10.636 & 2.52 \nl
 & & 14.0 & 1.91 & \phn$-3.84$ & 10.702 & 2.99 \nl
0.596 & ZAMS & 0.0 & 1.37 & \phn$-2.75$ & 10.563 & 2.18 \nl 
0.401 & ZAMS & 0.0 & 0.73 & \phn$-1.47$ & 10.412 & 1.61 \nl
\enddata
\tablenotetext{a}{in units of $10^{48}$ ergs.}
\end{deluxetable} 

\begin{deluxetable}{lcccccc} 
\tablecolumns{7}
\tablewidth{0pc}
\tablenum{2}
\tablecaption{Results of SPH Calculations} 
\tablehead{\colhead{Run\tablenotemark{a}} & \colhead{$M$} &
\colhead{$\frac{(a+b)}{2c}$} & \colhead{$E_{K}$\tablenotemark{b}} &
\colhead{$E_{I}$\tablenotemark{b}} & 
\colhead{$E_{G}$\tablenotemark{b}} & \colhead{$Y_{c}$}}
\startdata 
\multicolumn{7}{c}{analytic polytrope models} \nl
$n=\frac{3}{2}$ & 1.000 & 1.000 & 0.000 & 0.429 & $-0.857$ & \nl
$n = 3$ & 1.000 & 1.000 & 0.000 & 0.750 & $-1.500$ & \nl
$n = 4$ & 1.000 & 1.000 & 0.000 & 1.500 & $-3.000$ & \nl
\multicolumn{7}{c}{binary - binary collisions ($n=\frac{3}{2}$)} \nl
GH3 & 1.999 & 1.109 & 0.151 & 0.632 & $-2.132$ & 0.78 \nl
GH6 & 2.928 & 1.049 & 0.130 & 0.630 & $-3.213$ & 0.58 \nl
GH10 & 1.963 & 1.063 & 0.151 & 0.290 & $-1.655$ & 0.79 \nl
GH22 & 3.812 & 1.017 & 0.139 & 0.892 & $-4.329$ & 0.69 \nl
GH35 & 2.000 & 1.370 & 0.327 & 0.307 & $-1.868$ & 0.79 \nl
GH43 & 2.940 & 1.301 & 0.409 & 0.446 & $-2.768$ & 0.88 \nl
GH47 & 1.999 & 1.124 & 0.186 & 0.496 & $-1.985$ & 0.74 \nl
GH49 & 2.945 & 1.204 & 0.335 & 0.624 & $-3.283$ & 0.81 \nl
GH54 & 1.999 & 1.081 & 0.201 & 0.516 & $-2.020$ & 0.70 \nl
GH57 & 1.971 & 1.142 & 0.201 & 0.412 & $-1.846$ & 0.89 \nl
\multicolumn{7}{c}{single star interactions ($n = \frac{3}{2}$)} \nl
HCQ1 & 1.948 & 1.019 & 0.003 & 0.505 & $-1.922$ & 0.89 \nl
HCQ1BH & 1.964 & 1.009 & 0.004 & 0.609 & $-2.219$ & 0.96 \nl
GCQ1 & 1.986 & 1.104 & 0.159 & 0.162 & $-1.421$ & 0.62\tablenotemark{c} \nl
BMQ1 & 1.999 & 1.200 & 0.239 & 0.586 & $-2.023$ & 0.94 \nl
\multicolumn{7}{c}{single star interactions ($n = 3$)} \nl
HCQ1 & 1.861 & 1.020 & 0.002 & 1.733 & $-3.555$ & 0.83 \nl
HCQ1EM & 1.855 & 1.007 & 0.004 & 1.894 & $-3.814$ & 0.95 \nl
GCQ1 & 1.983 & 1.397 & 0.358 & 1.135 & $-3.045$ & 0.85 \nl
BMQ1 & 1.996 & 1.297 & 0.354 & 1.388 & $-3.557$ & 0.80 \nl
\enddata
\tablenotetext{a}{GH: Goodman \& Hernquist (1991) run; HC: head-on
collision; GC: grazing collision; BM: binary merger; EM: equal
mass SPH particles; BH: run mimicking Benz \& Hills (1987)}
\tablenotetext{b}{in units of $GM^{2}/R$.}
\tablenotetext{c}{maximum helium content $Y = 0.73$ occurs at $m / M = 0.02$}
\end{deluxetable} 

\begin{deluxetable}{ccccccc} 
\tablecolumns{7}
\tablewidth{0pc}
\tablenum{3}
\tablecaption{Results of Blue Straggler Stellar Evolution Calculations} 
\tablehead{\colhead{$M/M_{\odot}$} & \colhead{Stage\tablenotemark{a}}
&  \colhead{Age (yr)} & \colhead{$E_{I}$\tablenotemark{b}} &
\colhead{$E_{G}$\tablenotemark{b}} & \colhead{$\log R$ (cm)} &
\colhead{$n_{eff}$}}
\startdata 
\multicolumn{7}{c}{unmixed models} \nl
1.478 & ZA & 0.0 & 0.61 & \phn$-1.16$ & 11.719 & 2.15 \nl
 & R & $2.2 \times 10^{6}$ & 6.45 & $-12.98$ & 11.007 & 3.69 \nl
 & TO & $1.6 \times 10^{8}$ & 6.45 & $-12.97$ & 11.042 & 3.79 \nl
1.300 & ZA & 0.0 & 0.91 & \phn$-1.83$ & 11.403 & 2.11 \nl
 & R & $5.6 \times 10^{6}$ & 5.45 & $-10.98$ & 10.879 & 3.39 \nl
 & TO & $7.8 \times 10^{8}$ & 5.45 & $-10.97$ & 10.957 & 3.66 \nl
1.071 & ZA & 0.0 & 0.59 & \phn$-1.18$ & 11.358 & 1.63 \nl
 & R & $1.4 \times 10^{7}$ & 3.84 & \phn$-7.75$ & 10.815 & 3.21 \nl
 & TO & $3.65 \times 10^{9}$ & 4.14 & \phn$-8.32$ & 10.939 & 3.74 \nl
\multicolumn{7}{c}{completely mixed models} \nl
1.478 & ZA & 0.0 & 6.97 & $-14.06$ & 10.802 & 3.06 \nl
 & TO & $6.5 \times 10^{8}$ & 6.88 & $-13.83$ & 10.918 & 3.49 \nl
1.300 & ZA & 0.0 & 5.44 & $-10.99$ & 10.809 & 3.11 \nl
 & TO & $1.57 \times 10^{9}$ & 5.59 & \phn$-11.25$ & 10.934 & 3.62 \nl
1.071 & ZA & 0.0 & 3.81 & \phn$-7.70$ & 10.806 & 3.12 \nl
 & TO & $4.07 \times 10^{9}$ & 4.15 & \phn$-8.34$ & 10.939 & 3.75 \nl
\enddata
\tablenotetext{a}{ZA: zero-age (initial) model; R: relaxed model; TO: turnoff model}
\tablenotetext{b}{in units of $10^{48}$ ergs.}
\end{deluxetable} 


\begin{references}

\reference{al} Alexander, D. R. 1996, private communication

\reference{ats} Anthony-Twarog, B. J., Twarog, B. A., \& Suntzeff, N.
B. 1992, \aj, 103, 1264

\reference{aol} Auri\`{e}re, M., Ortolani, S., \& Lauzeral, C. 1990,
\nat, 344, 638

\reference{b} Bailyn, C. D. 1992, \apj, 392, 519

\reference{b} Bailyn, C. D. 1995, \araa, 33, 133

\reference{bp} Bailyn, C. D. \& Pinsonneault, M. C. 1995, \apj, 439, 705

\reference{bha} Benz, W. \& Hills, J. G. 1987, \apj, 323, 614

\reference{bhb} Benz, W. \& Hills, J. G. 1992, \apj, 389, 546

\reference{cpc} Carretta, E. \& Gratton, R. G. 1996a, \aaps, in press

\reference{cg} Carretta, E. \& Gratton, R. G. 1996b, in ASP Conf. Ser.
92, Formation of the Galactic Halo\ldots Inside and Out, ed. H.
Morrison and A. Sarajedini, (San Francisco: ASP), 363

\reference{css} Chieffi, A., Straniero, O., \& Salaris, M. 1991, in
ASP Conf. Ser. 13, The Formation and Evolution of Star Clusters, ed.
K. Janes (San Francisco: ASP), 219

\reference{d} Dorman, B. 1992a, \apjs, 80, 701

\reference{d} Dorman, B. 1992b, \apjs, 81, 221

\reference{egg} Eggleton, P. P. 1991, private communication

\reference{eff} Eggleton, P. P., Faulkner, J., \& Flannery, B. P.
1973, \aap, 23, 325

\reference{fff} Ferraro, F. R., Fusi Pecci, F., Cacciari, C., Corsi,
C., Buonanno, R., Fahlman, G. G., \& Richer, H. B. 1993, \aj, 106, 2324

\reference{fpf} Fusi Pecci, F., Ferraro, F. R., Corsi, C. E., Cacciari,
C., \& Buonanno, R. 1992, \aj, 104, 1831

\reference{gh} Goodman, J. \& Hernquist, L. 1991, \apj, 378, 637

\reference{gqo} Gratton, R. G., Quarta, M. \& Ortolani, S. 1986,
\aap, 169, 208

\reference{her} Hernquist, L. 1993, \apj, 404, 717

\reference{hk} Hernquist, L. \& Katz, N. 1989, \apjs, 70, 419

\reference{hnrf} Hodder, P. J. C., Nemec, J. M., Richer, H. B., \&
Fahlman, G. G. 1992, \aj, 103, 460

\reference{hrev} Hut, P., et al. 1992, \pasp, 104, 981

\reference{irw} Iglesias, C. A., Rogers, F. J. \& Wilson, B. G. 1992,
\apj, 397, 717

\reference{j} J$\o$rgensen, H. E. 1982, \aap, 108, 99

\reference{jh} J$\o$rgensen, H. E. \& Hansen, L. 1984, \aap, 133, 165

\reference{kk} Kaluzny, J. \& Krzeminski, W. 1993, \mnras, 264, 785

\reference{lar} Lai, D., Rasio, F. A., \& Shapiro, S. L. 1993, \apj,
412, 593

\reference{lms} Lambert, D. L., McWilliam, A., \& Smith, V. V. 1992,
\apj, 386, 685

\reference{lac} Lauzeral, C., Auri\`{e}re, M., \& Coupinot, G. 1993,
\aap, 274, 214

\reference{loa} Lauzeral, C., Ortolani, S., Auri\`{e}re, M., \& Melnick,
J. 1992, \aap, 262, 63

\reference{ldz} Lee, Y.-W., Demarque, P., \& Zinn, R. 1990, \apj, 350, 155

\reference{ll} Leonard, P. J. T. \& Livio, M. 1995, \apj, 447, L121

\reference{lrs} Lombardi, J. C., Rasio, F. A., \& Shapiro, S. L. 1995,
\apj, 445, L117

\reference{mhn1} Mateo, M., Harris, H. C., Nemec, J., Olszewski, E.,
\& Schombert, J. 1988, \baas, 20, 717

\reference{mhn} Mateo, M., Harris, H. C., Nemec, J., \& Olszewski, E.
W. 1990, \aj, 100,469

\reference{m} Mathys, G. 1991, \aap, 245, 467

\reference{m} Monaghan, J. J. 1992, \araa, 30, 543

\reference{ml} Monaghan, J. J. \& Lattanzio, J. C. 1985, \aap, 149, 135

\reference{nc} Nemec, J. M. \& Cohen, J. G. 1989, \apj, 336, 780

\reference{nmb} Nemec, J. M., Mateo, M., Burke, M., \& Olszewski, E.
W. 1995, \aj, 110, 1186

\reference{njl} Niss, B., J$\o$rgensen, H. E., \& Lausten, S. 1978, \aaps, 32, 387

\reference{p} Paresce, F., et al. 1991, \nat, 352, 297

\reference{dm} Peterson, C. J. 1993, in ASP Conf. Ser. 50, Structure and
Dynamics of Globular Clusters, ed. S. G. Djorgovski and G. Meylan (San
Francisco: ASP), 337

\reference{pcl} Peterson, R. C., Carney, B. W., \& Latham, D. W. 1984,
\apj, 279, 237

\reference{p} Pilachowski, C. A. 1984, \apj, 281, 614

\reference{poo} Pilachowski, C. A., Olszewski, E. W. \& Odell, A. 1983,
\pasp, 95, 713

\reference{pbd} Proctor Sills, A., Bailyn, C. D., \& Demarque, P.
1995, \apj, 455, L163

\reference{rs} Rasio, F. A. \& Shapiro, S. L. 1995, 438, 887

\reference{fr} Rogers, F. J., 1996, private communication

\reference{rb} Rubenstein, E. P. \& Bailyn, C. D. 1996, \aj, 111, 260

\reference{op} Seaton, M. J., Yan, Y., Mihalas, D., \& Pradhan, A. K.
1994, \mnras, 266, 805

\reference{sm} Smith, H. A. 1984, \apj, 281, 148

\reference{s} Stryker, L. L. 1993, \pasp, 105, 1081

\reference{tl} Taam, R. E. \& Lin, D. N. C. 1992, \apj, 390, 440

\reference{v} VandenBerg, D. A. 1992, \apj, 391, 685

\reference{wkm} Weiss, A., Keady, J. J., \& Magee, N. H. 1990, Atomic
Data and Nuclear Data Tables, 45, 209

\reference{ym} Yan, L. \& Mateo, M. 1994, \aj, 108, 1810

\reference{yn} Yan, L. \& Reid, I. N. 1996, \mnras, 279, 751

\reference{ygbs} Yanny, B., Guhathakurta, P., Bahcall, J. N., \&
Schneider, D. P. 1994a, \aj, 107, 1745

\reference{ygbs} Yanny, B., Guhathakurta, P.,
Schneider, D. P, \& Bahcall, J. N. 1994b, \apjl, 435, 59

\reference{zw} Zinn, R. \& West, M. J. 1984, \apjs, 55, 45

\end{references}
\end{document}